# A model to assess customer alignment through customer experience concepts
## (Preprint version)

Leonardo Muñoz[1] and Oscar Avila[1]

[1] Department of Systems and Computing Engineering, School of Engineering,
Universidad de los Andes, Bogotá, Colombia
`{l.munozm, oj.avila}@uniandes.edu.co`

**Abstract.** Business and Information Technology Alignment (BITA) has been one of the main concerns of IT and Business executives and directors due to its importance to overall company performance, especially today in the age of digital transformation. For BITA has been developed several models which in general has focused in the implementation of alignment strategies for the internal operation of the organizations and in the measurement of this internal alignment, but, there is still a big gap in measurement models of the alignment with the external environment of the organizations. In this paper is presented the design and application of a maturity measurement model for BITA with the customers, where the customers are actors of the external environment of the companies. The proposed model involves evaluation criteria and business practices which the companies ideally do for improve the relationship with their customers.

## 1    Introduction

The organizational context is characterized today by fast and unexpected changes what force organizations to adapt their internal operation and relationship with the environment in order to be according to new issues, needs and requirements. Following this operational and strategical transformations, the technology has been included as an enabler of many of the new business capabilities and, in this way, it has taken an important place into any business model.

In this context, Business and Information Technology Alignment (BITA) is a key process for the performance of the organizations in a fast changing and highly demanding market [1] [2]. This process must be approached from three different levels: (i) the internal level that consists in aligning IT services and business processes; (ii) the alignment with environment that concerns the alignment of the organization with external actors, issues and needs; (iii) the alignment with uncertain evolutions that relates aligning the organization with external and internal overcoming events and changes [3].

Due to the importance of BITA in last 20 years, its measurement has becoming important in order to get insight about the current maturity level of the organizational capabilities allowing companies to align IT and Business domains. Such picture makes possible to identify and correct gaps in the progress made in both domains and to identify the impact of organizational and digital transformation initiatives.

A literature review [4], looking for contributions in BITA measurement at the three beforementioned alignment levels, shown that the wide research work is focused on the measurement of the internal alignment level with great advances and many approaches. This analysis showed as well that there is an absence of measurement models for the alignment with the external environment and future evolutions. However, these two levels are increasingly becoming relevant in the business world, because of the emerging importance of actors, such as customers, competitors and suppliers that generate pressure for change on the environment in which companies compete.

Keeping in mind this gap, this work focuses on the alignment with the external environment and specifically with customers, which have become more exigent and digitally connected and have increased their negotiation power. To this end, we adopt additional concepts such as digital transformation (DT) [5] because to address this new challenge, companies are today trying to improve the customer experience by profiting of new disruptive technologies and implementing digital transformation projects. In this work, DT frameworks are taken as a basis, in conjunction with the existing measurement models in order to propose a maturity model for measuring the strategic alignment with customers.

This article is structured as follows: Chapter 2 presents the most relevant works of the state of the art as well as the identified gaps, chapter 3 presents the approach of the new model for measuring the maturity of the alignment with customers, chapter 4 presents the case study under which the proposed measurement model was applied and by Last, chapter 5 presents the conclusions and future work.

## 2    State of the Art

In this section we present main findings and gaps into the state of the art about BITA measurement, published in [4]. Also are presented selected relevant works about BITA focused on the customer experience, what is a key pillar for the develop of the model presented in this article.

The purpose of the state of the art is to evaluate the last 15 years of research literature on BITA measurement methods and approaches. For the evaluation of this set of articles, the proposed framework in [3] was adopted as a basis for developing a personalized evaluation framework. To classify the BITA approaches, the Strategic Alignment Model (SAM) [6] proposed by Henderson and Venkatraman was adopted as one of the basic criteria to evaluate the scope of the reviewed works. Based on this, a characterization framework was defined to review the current status of BITA measurement methods and approaches.

**Gaps identified in the measurement of business and IT alignment.**

According to the review of the current state of development regarding the topic of business and IT alignment measurement, it is found that there is extensive work focused on the internal-classical alignment for the performance improvement of the companies' processes, involving more the IT areas within the business domain. Despite this, it is evident the lack of development of the research in alignment measurement of two remaining levels, first, the alignment with the future and the uncertain evolutions, and second, the alignment with the external environment of the company.

The lack of models to measure the business and IT alignment with the external environment is very important because traditional and modern companies converge and compete in a highly changing market. In addition to this, the new business models increasingly include the outsourcing of services and process, and temporary partnerships into their operation, which increases the need to enable new capabilities to the outside of the business, for the relationship with the external actors, driven by technological tools that the IT department should enable and manage.

For approach the problem of lack of alignment measurement models with the external environment, it is necessary to delimit the scope of this work. For this reason and considering that for many companies today their clients are a key point in their strategies [7], this work focuses on the alignment with the external environment and specifically with the customers as the main actors in the business environment.

**Digital Transformation of Customer Experience**

According to this approach of alignment with customers as an external actor, is fundamental to consider the digital transformation (DT) advances as a base of development for the proposed maturity model, because the DT generally includes the customer experience transformation as a pillar in their frameworks as in [8], [9], [5], and specifically is presented in the DT framework by Capgemini consulting and MIT [5]. According to [10] the new customers are generating new and faster requirements that must be solved by the digitalization of the customer experience, using the technological tools to: understand the customer and external context, improve and maximize the physical experience with digital tools and focus on smart investments in digital channels.

An additional key point in the improvement of the alignment with the customers by doing a DT in the companies, is to focus on generating value to the customers through their lifecycle, as is described in [11], what suggests the companies must to know better their customer's needs, tastes and tendencies to be agile and flexible in the decision making process. It reinforces the idea of technology as the main tool in the business models to enable new capabilities in the companies to align to their customers and improve the performance.

## 3 Design and development of the Maturity Measurement Model

Considering that alignment with customers is focused on close the gaps between the companies' strategies and the preferences, needs and tastes of clients, it is necessary to take in consideration all the customer lifecycle. This lifecycle is composed by discovering of customer needs, attracting the customer, following the purchase process and finally supporting post-sale services [12]. For aligning the customer lifecycle with the company strategy and processes, it is very important to focus on the customer experience through the whole process. The better customer experience, the more client attraction, more effective sales and more faithful clients, what means the possibility of new clients attracted by better feelings and opinions of the current clients [13].

By this way, for the definition of the maturity measurement criteria, is taken as a reference of development the "Customer Experience" pillar of the Digital Transformation Framework [5], which is focused on the transformation of the relationships with the users and clients through digital tools and capabilities. These measurement criteria include all the customer lifecycle from the pre-sale stage, then sales process and finally post-sale services.

### 3.1 Model Proposal Methodology

For structuring the maturity model, we propose the next steps:

**Structure Definition:** The mode was designed using a two-dimensional structure (measurement criteria and maturity levels) which are very common in maturity models designed in the academy and the industry [14]. Also was considered the SAMM maturity model by Luftman [15], adopting from this the structure of "business practices" that will be evaluated into each measurement criteria.
**Criteria and business practices definition:** for this step was taken as a base of development the DT Framework of Capgemini, and specifically was adopted the customer experience pillar from it. From that pillar was taken their elements as the measurement criteria and business practices, and these was also complemented with concepts from the state of the art.
**Measurement scale definition:** for the measurement scale was adopted the same Likert scale used in the SAMM model, composed by 5 maturity levels which will be evaluated for each criteria and business practice.
**Evaluation references definition:** according to the SAMM structure adopted, for each business practice is proposed a reference state description for each maturity level, what works as comparative reference for the evaluation process. These reference states are proposed under the experience of several experts were consulted during the model development.

### 3.2 Measurement Criteria.

**Customer understanding:** the objective of this criterion is to evaluate the maturity of the customers information obtaining processes, to improve the understanding of their preferences, behaviors and tendencies in the market, and enable the customization of the products and services offers.

This criterion groups the following practices that must be considered to assess the level of customer understanding:

- *Customer segmentation based on information analysis:* This practice assesses the use of technological tools such as databases and business intelligence applications to classify clients according to information obtained from internal and external sources.
- *Customer sentiments analysis:* aims to evaluate the ability to obtain and analyze information on customer sentiments and opinions about products, services or company image through technological tools.
- *Behavior and tastes analysis of potential clients:* The ability to obtain information on the profile of current and prospective clients regarding their tastes, behaviors and preferences through technological tools is evaluated.
- *Management of current customer base with computer systems:* Within this practice the obtaining, retention and management of the information of the company's current customers is evaluated through computer tools, to simplify the processes involved, speeding up access for decision making.
- *Integration of customer information sources:* This practice evaluates the level of updating, management and consolidation of multiple sources of customer information through technological tools, to have centralized data warehouses for advanced analysis of data.

**Marketing and sales process:** This criterion aim to evaluate the use of different technological tools to transform the net revenue channels of companies, with their associated processes. In addition to these processes, the use of attraction and customer retention tools is also measured.

The practices that are grouped within this criterion are:

- *Use of electronic sales channels:* This practice evaluates the level of implementation of digital sales tools or digital sales assistance, as well as their integration among the different channels.
- *Use of electronic marketing channels:* This practice evaluates the implementation and proper use of multiple electronic marketing channels.
- *Predictive marketing implementation*: In this practice, the level of use of business intelligence in marketing processes is evaluated.
- *Digitization of sales operative processes towards customers:* Within this practice, the use of technological assistance tools or automation of the sales process is evaluated to make it much more agile and effective for the clients.
- *Mobility in the sales process:* In this practice the objective is to evaluate the level of availability of sales tools in different mobile media, which facilitate the relationship with the customer in the sales process.
- *Visibility of sales processes to the client:* For this practice the level of visibility and control of the sales process allowed to the client is evaluated, to generate a personalized and secure experience towards this.

**Customer service:** This criterion aims to evaluate the improvement in speed and effectiveness of solution of post-sale requirements of customers through the new digital relationship channels.

The practices that are grouped and evaluated in this criterion are:

- *Use of digital channels for customer service:* This practice aims to evaluate the use of different digital channels of communication with the customer to respond to their requirements in a flexible and agile way.
- *Coherence between the communication channels used with clients:* In this practice, the level of coherence that exists between the customer service channels implemented is evaluated.
- *Implementation of simple and agile service technology tools:* This practice evaluates the level of simplicity and efficiency of the service tools for the customer. It is not enough to implement technological tools, but they must be friendly and agile to guarantee the best possible experience.
- *High availability of digital service channels:* Within this practice, the continuity and availability of service tools for the customer is evaluated.
- *Use of self-service tools for requirements:* The existence of platforms that allow customers to have control over the solutions to their requirements is evaluated and, in the best case, they can obtain solutions on their own in a simple and agile way.
- *Service experience feedback channels:* This practice evaluates the availability of feedback channels for the user experience in relation to customer service.

Having already defined the measurement criteria and their associated practice groups, it is proposed to evaluate them under 5 maturity levels defined from the Likert scale in the SAMM levels and adapted to the particularities of the customer experience area.

1. *Initial / Process Ad Hoc.* - Practice not implemented or not aligned between IT and business.
2. *Committed process.* - The organization has plans to implement and improve the practice.
3. *Focused and stabilized process.* - The practice is established but external alignment is still lacking from IT and business.
4. *Improved / Managed Process.* - Within the practice IT is conceived as an element of value within the business strategies towards the external environment of clients.
5. *Optimized Process.* - The practice has been implemented in its entirety and is flexible in the face of changes in the client and organization market.

With these general levels, specific states are established for each of the practices established within the criteria, which will be the response guide for the evaluation applied to the company. See Table 1.

*Table 1. Maturity measurement framework for Business and IT Alignment with the Clients.*

| Crit. | Practice | Level 1 | Level 2 | Level 3 | Level 4 | Level 5 |
|---|---|---|---|---|---|---|
| Understanding of the client | **Segmentation of clients based on information analysis.** | There is no segmentation of the customer base | Clients inaccurately segmented from incomplete information, however, there are plans to improve the information sources and analysis tools | Clients segmented based on local data analysis. Analysis tools and obsolete or limited sources. | Clients segmented based on local data analysis. CRM tools managed by the IT area are used | Clients segmented based on analysis of local and external data. CRM tools and business intelligence are used managed by the IT area |
| | **Analysis of customer sentiments.** | There are no tools or data sources for customer sentiment analysis. | Manual monitoring and analysis are carried out from a single source of information. | Monitoring and semi-automatic analysis of several sources with isolated tools for each source. | Semi-automatic monitoring and analysis is performed consolidating data from multiple sources. | Automatic monitoring and analysis are performed consolidating data from multiple sources, using filters and artificial intelligence. |
| | **Analysis of the behavior and tastes of potential clients.** | There are no tools or data sources for analysis of behavior and preferences. | Manual monitoring and analysis are done with local and limited data sources. | Semi-automatic monitoring and analysis is done with some data capture tools in web portals and social networks. | Automatic monitoring and analysis are done with centralized tools and multiple web sources of information. | Automatic monitoring and analysis are done with centralized and intelligent tools and multiple web and IoT sources. |
| | **Management of current customer base with computer systems.** | There is no database of current customers | Clients are managed with an outdated database. | Clients are managed in a local database with occasional updates. | Clients are managed with a CRM system with little adoption and inadequate use. | Clients are managed with an updated CRM system in real time through mobile and local tools. |
| | **Integration of information sources of current customers and prospects.** | There is no integration strategy for information sources. | There are isolated sources of information with plans for future integration. | The information sources are partially integrated and are updated with low frequency. | There is a central data bank that is frequently updated with information from a limited number of sources. | There is a central data bank that is updated in real time with customer information through multiple channels. |
| Marketing and sale process | **Use of electronic sales channels.** | There are no electronic sales channels. | There is a catalog of products within the web portal | The web portal implements an online sales platform with some products and a complex process. | The web portal implements an online sales platform of its complete portfolio and with a simple process. Managed by the IT and sales area. | There are multiple electronic sales channels through social networks and the web portal, managed by the IT and sales area. |
| | **Use of electronic marketing channels** | There are no electronic marketing channels | There are plans to implement electronic marketing media, emails and web pages. | The email is used to send marketing material to customers. There is no monitoring of its effectiveness. | Email and social networks are used as marketing channels and there is manual monitoring of effectiveness. | Consolidation tools of electronic marketing channels are used and full monitoring of the effectiveness of the campaigns is made. |
| | **Implementation of predictive marketing** | There is no predictive marketing | Marketing based on limited customer information | Marketing based on full local customer information | Marketing based on trends of local and external customer information | Marketing based on predictive models in artificial intelligence tools and with local and external customer information |
| | **Digitalization of operative sales processes towards clients.** | Completely manual sales processes | Manual sales process with digitization plan in progress | Partially digitized sales process with electronic payment | Digitized sales process with paper support documents | Sales process completely digital and connected to the company's information systems. |
| | **Mobility in the sale process.** | There are no sales tools accessible through mobile devices | There are electronic sales tools with projection of implementation of mobile access | There are sales tools with the possibility of mobile access, but without implementation of their use | Sales tools have mobile access implemented | Sales tools have mobile access implemented and real-time connection to information systems |
| | **Visibility of sales processes to the client.** | The sales process is not visible to the customer | The sales process is partially visible to the customer | The sales process is visible to the customer in its entirety without granting control to the client. (tracking) | The client has visibility and partial control of the sales process. (Ex Times and delivery places) | The sales process is visible and customizable for the client. (Eg. Shopping channels, offers, times, deliveries) |
| Customer service | **Use of digital channels for customer service.** | There are no digital channels for customer service | There is an implementation plan of the customer service channel by email | Customer service is provided through the contact section of the web portal | Customer service is provided through different digital channels such as social networks and web portal | There is a centralized management platform for customer service through social networks and a web portal with information available to customers. |
| | **Coherence between the communication channels used with clients.** | The communication channels with customers do not share information between them | There is a plan to integrate information from traditional communication channels such as email and web portal | Traditional communication channels share information with low frequency | The traditional communication channels are integrated and coherent by the management of the IT area. The new channels have partial integration. | Traditional communication channels are integrated and coherent with new channels such as social networks. The IT area has visibility and total control. |
| | **Implementation of simple and agile service technology tools.** | There are no technological tools for customer service | There are plans to implement traditional technological tools for customer service | There are obsolete technological tools for customer service with high levels of complexity | There are agile customer service technology tools with multichannel integration | There are advanced customer service tools and with some level of autonomy |
| | **High availability of digital service channels.** | Customer service channels do not have | There is a project to implement high availability strategies in service channels | The main service channel has high availability, but is not implemented for all customer | Traditional customer service channels have high availability. No new service channels | It has high availability for traditional service channels and is integrated with channels in |

| | | | | | |
|---|---|---|---|---|---|
| | high availability strategies | | service channels | are contemplated. | the cloud as social networks. |
| **Use of self-service tools of requirements.** | No self-service requirements tools are implemented | The requirements request tools have a projection for the implementation of self-service modules. | Requirement request tools allow self-service partially for clients. | There is a complete self-service platform for customer requirements with occasional assistance by the service team. | There is a complete self-service platform for customer requirements assisted by intelligent systems. Minimum human assistance. |
| **Feedback channels of service experience.** | Service experience feedback channels are not implemented. | There is a basic survey on unconsolidated paper. There is a plan to implement digital channels. | A service comment box is available through a web portal or email. | The service tools implement digital feedback channels associated with the requirements. (online forms) | There is multi-channelity for the feedback of the service experience. The channels are integrated and relate to the requirements of each client. |

### 3.3 Methodology of application of the evaluation framework:

For the application of this model to measure the maturity of the alignment of IT and business with customers, it is necessary to define an evaluation group made up of executives or business and IT managers, who are in charge of answering and discussing the levels of maturity that adjust to the current state of the company in each of the criteria and business practices, and most importantly, define the gaps found and the possible actions to be taken from the results [15].

The specific steps for the application of the measurement model take as reference the methodology of the SAMM model [16] by Luftman, that includes the following 4 steps:

1. *To conform the evaluation team:* create a team of IT and business executives for evaluation. The number of executives varies depending on the size of the company and whether a business unit or the entire organization is evaluated.
2. *To gather information:* The defined team must evaluate each of the business practices of each of the criteria. This can be done in three ways:
   a. In joint work of the team.
   b. Responding to surveys by each of the members and meeting to discuss and consolidate results. For this, an individual online evaluation format is used.
   c. Combining the first two methods in case some of the members have difficulties for meetings.
3. *To decide the individual scores:* The team must reach an agreement to assign a score to each of the practices evaluated, highlighting the gaps found and the possible steps to be followed to solve them. The reference score for each of the criteria will be the average of the scores of the practices that it groups.
4. *To decide the overall alignment score:* The team must achieve a consensus of the total alignment score assigned to the current state of the company. The average scores of each criterion will guide this consensus, but the team can adjust the total score if it considers that certain practices have more or less weight within the company than others, according to their industry context.
5. *To present an executive report:* After consolidating the partial scores and the total score, it is recommended to prepare an executive summary of the evaluation for the board of directors, which includes the levels obtained, gaps found and possible improvement strategies. It is proposed that you use a general evaluation consolidation format as a reference.

## 4  Case Study

This section presents the results of the case study of one of the three anonymous companies from different industries, sizes and business models where the model was applied, it includes the application of the measurement model, following the methodology proposed in the previous chapter. The participating companies have operations in Colombia and Latin America, but they are not a significant sample, so it is not intended to characterize the market, but only to document the experience of applying the model in a real business environment.

The first part of the case study is the profiling of company. The second part of the case study presents the results obtained from the application of the measurement model in the selected company and the analysis of the results.

**Profile of participating company:** Below, the profile of the company selected, from the three participating companies in the case study, are presented:

*Industrial sector:* Technology and services
*Number of employees:* 50 to 200
*Description of economic activity:* Business unit of a multinational technology and services company, responsible for carrying out the pre-sale and sale of business technology solutions, serving corporate clients throughout the Latin American region. This business unit has employees in several countries in the region.
*Approximate number of customers:* 20000

### 4.1  Results of application of the measurement model

According to the methodology of application of the measurement model, there are two stages in obtaining results, the first is the average results of the evaluations, made by the members of the evaluation team individually, and the second is the results consolidated under consensus of the entire evaluation team with the identification of gaps and possible future strategies. In this article only will be presented the consolidated results.

**Results by consensus of the evaluation team:**
After obtaining the results of the evaluations carried out individually by each member of the evaluation teams, each company proceeded to hold an evaluation meeting in which the levels of maturity obtained in each criterion and evaluated practice were discussed and adjusted.

Below in table 2, the consolidated results of the company are presented with the weights for each criterion, as well as the most important gaps that were identified within the evaluation process.

*Table 2. Results by consensus of evaluating team of Company.*

| Criterion | Practice | Average level | Weighting in% | Average by criterion |
|---|---|---|---|---|
| Understanding of the client | Segmentation of clients based on information analysis. | 4,2 | 25% | 3,54 |
| | Analysis of customer sentiments. | 2,2 | 0% | |
| | Analysis of the behavior and tastes of potential clients. | 2,8 | 25% | |
| | Management of current customer base with computer systems. | 3,8 | 25% | |
| | Integration of information sources of current and prospective clients. | 3,3 | 25% | |
| Sales process | Use of electronic sales channels. | 3,3 | 16,67% | 3,3 |
| | Use of electronic marketing channels. | 4,0 | 16,67% | |
| | Predictive marketing implementation | 2,7 | 16,67% | |
| | Digitization of sales operative processes towards clients. | 3,5 | 16,67% | |
| | Mobility in the sale process. | 3,7 | 16,67% | |
| | Visibility of sales processes to the client. | 2,8 | 16,67% | |
| Customer service | Use of digital channels for customer service. | 4,0 | 16,67% | 3,3 |
| | Coherence between the communication channels used with clients. | 2,8 | 16,67% | |
| | Implementation of simple and agile service tools. | 3,5 | 16,67% | |
| | High availability of digital service channels. | 3,2 | 16,67% | |
| | Use of self-service tools of requirements. | 4,0 | 16,67% | |
| | Service experience feedback channels. | 2,5 | 16,67% | |
| General level: | 3,4 | | | |

**Main gaps identified:**
In the understanding of the client criterion, it was identified that, although there are advanced CRM tools for information management, full and adequate use of all the functionalities available for the relationship with the clients are not implemented. Problems of information quality and source integration are also identified.

In the sales and marketing processes, it was identified that, despite being a business to business model, it is possible to have more agile technological tools for the processes of placing and tracking purchase orders for the final customers. Mobile tools for sales exist, but their full and adequate use has not been implemented due to lack of management from the managerial level. There are plans to implement artificial intelligence within the marketing processes, but it is not yet available for use within the business unit.

In the post-sales customer service, it was identified that technological tools lack self-service options and there are problems of availability of resources to respond quickly to support cases. Nor is there a user experience feedback channel. The implementation of artificial intelligence was identified as an improvement option to assist customers in the post-sales support processes.

**Analysis of the results**
The average scores obtained from the individual evaluations made by the team members were maintained. Regarding the weight of the evaluated practices, relevance to customer sentiment analysis was removed due to, within a business model with enterprise clients, it is not of vital importance for its operation. The remaining practices maintained an equal weight within the evaluation.

As a business unit of a multinational technology company with a presence throughout Latin America and a long history, the operating characteristics of this evaluated unit are quite mature and with a very good adoption of the technology within its business strategy. Even so, there are significant gaps in the alignment of IT and business with customers as observed in the results obtained, for this reason it is at a level of maturity 3.4 that is somewhat higher than level 3 of focused and stabilized processes. The unit has practices of using technological tools with a higher level in understanding of the client and in the sales process, and this positions the company, in general, in an advanced range of maturity.

Among the proposals for improvement for this company are the evaluation and projection of the use of new technologies that are cutting edge in the business environment such as artificial intelligence, as well as the implementation of information systems integration processes. This would help reach a level 4 maturity of the alignment of IT and business with customers.

## 5      Conclusions and future work

For improving strategic alignment with customers, today companies are improving the relationships and the experience of the customers by making many efforts and investing resources in the adoption of new technologies within their business models as capacity enablers, this is better known as "Digital Transformation" (TD).

A model was developed to measure the maturity of the alignment of IT and business with customers of a qualitative nature, based on the development of the SAMM model. From the application of this model in three companies of different profiles, it was concluded that the maturity of the alignment with customers is a subject that still has a relatively low level of maturity due to the little development it has had in recent years.

From this, it is concluded that, although the level of maturity of the alignment with customers is low due to the lack of formality in the approach, even so, companies have a basic level of alignment and are aware of its importance.

It is also observed that, although the measurement model has reference states at each level for each of the evaluated practices, they cannot cover all the business models and industries in a generic way due to the high complexity that each context implies. Additionally, the qualitative nature of the maturity measurement model allows a high level of subjectivity in the evaluation, which is sought to mitigate with the discussion meeting of the evaluation team.

Finally, it can be concluded that to make the evaluation process more precise, it is necessary to include an external business consultant or business architect that helps generate a more neutral environment in the evaluation process, further reducing the subjectivity of the application of the Maturity measurement model.

**Future work:**

Based on this work, the following important topics were identified for future work on the area of research in the measurement of IT and business alignment:

Approach of measurement models of IT alignment and business with the external environment of the organizations, focused on the actors that make up that external environment, in addition to the clients. Example of this is the alignment with suppliers, business associates, competitors, etc.

Validation and adjustment of the measurement model of alignment with customers, applying it in different industries and obtaining feedback to adjust the criteria and practices evaluated. It is also possible to propose methodological complements to the measurement model to avoid subjectivity and make it more faithful in the results.

Approach of measurement models for the alignment of IT and business with future evolutions, through analysis and predictive tools that allow to anticipate changes in the internal and external environment of the companies.

## References

[1] J. Luftman and B. Derksen, "Key Issues for IT Executives 2012: Doing more with less," *MIS Quarterly Executive,* pp. 207-218, 2012.

[2] J. Luftman, B. Derksen, R. Dwivedi, M. Santana, H. S. Zadeh and E. Rigoni, "Influential IT management trends: an international study," *Journal of Information Technology,* vol. 30, pp. 293-305, 2015.

[3] O. Avila, V. Goepp and F. KIEFER, "UNDERSTANDING AND CLASSIFYING INFORMATION SYSTEM ALIGNMENT APPROACHES," *The Journal of Computer Information Systems,* vol. 50, no. 1, pp. 2-14, 2009.

[4] L. Muñoz and O. Ávila, "Business and Information Technology Alignment Measurement - A Recent Literature Review," in *Business Information Systems Workshops. BIS 2018. Lecture Notes in Business Information Processing*, 1 ed., vol. 339, Cham, Springer International Publishing, 2019, pp. 112-123.

[5] MIT Center for Digital Business and Capgemini Consulting, "Digital Transformation: A roadmap for Billion-Dollar Organizations," MIT Center for Digital Business and Capgemini Consulting, 2011.

[6] J. C. Henderson and N. Venkatraman, "Strategic alignment: Leveraging information technology for transforming organizations," *IBM Systems Journal,* vol. 32, no. 1, pp. 4-16, 1993.

[7] MIT Sloan Management Review and Google, "RESEARCH REPORT Leading With Next-Generation Key Performance Indicators," Massachusetts Institute of Technology, 2018.

[8] J. Reis, M. Amorim, N. Melão and P. Matos, "Digital Transformation: A Literature Review and Guidelines for Future Research," *Trends and Advances in Information Systems and Technologies. WorldCIST'18 2018. Advances in Intelligent Systems and Computing,* vol. 745, p. 411–421, 2018.

[9] Harvard Business Review Analytic Services, "THE DIGITAL TRANSFORMATION OF BUSINESS," Harvard Business School Publishing., 2015.

[10] Bonnet, Didier; Buvat , Jerome; KVJ, Subrahmanyam ; Digital Transformation Research Institute, "Rewired: Crafting a Compelling Customer Experience," *Digital Transformation Review,* no. 6, 2014.

[11] S. Vandermerwe, "How Increasing Value to Customers Improves Business Results," *MIT Sloan Management Review,* vol. 42, no. 1, pp. 27 - 37, 2000.

[12] F. Buttle, Customer Relationship Management, Concepts and Tools, Oxford: Elsevier Butterworth-Heinemann, 2004.

[13] B. Morgan, "Forbes - Breathing New Life Into The Customer Lifecycle," 24 Abril 2017. [Online]. Available: https://www.forbes.com/sites/blakemorgan/2017/04/24/breathing-new-life-into-the-customer-lifecycle/#4f0d19014294. [Accessed 19 Septiembre 2018].

[14] J. Becker, R. Knackstedt and J. Pöppelbuß, "Developing Maturity Models for IT Management," *Business & Information Systems Engineering,* vol. 1, no. 3, pp. 213-222, 2009.

[15] J. Luftman, "Strategic Alignment Maturity," in *Handbook on Business Process Management 2*, Berlin Heidelberg, Springer-Verlag, 2015, pp. 5 - 43.

[16] J. Luftman, "Assessing It/Business Alignment," *Information Systems Management,* vol. 20, no. 4, pp. 9-15, 2003.